\documentclass{PoS}

\usepackage{epsfig}
\usepackage{latexsym}
\usepackage{amsmath}
\usepackage{amssymb}

\renewcommand{\paragraph}[1]{\vspace{1cm}\par{\Large\bf #1}\par}

\newcommand{\be}{\begin{equation}}
\newcommand{\ee}{\end{equation}}
\newcommand{\QchiPT}{Q$\chi$PT}

\title{Light quark electromagnetic structure of baryons}

\ShortTitle{Light quark electromagnetic structure of baryons}

\author{
S.~Boinepalli%
J.~N.~Hedditch,$^{a}$
B.~G.~Lasscock,$^{a}$
D.~B.~Leinweber,$^{a}$
\speaker{A.~G.~Williams},$^{a}$
J.~M.~Zanotti$^{b}$
and J.~B.~Zhang$^{c}$
\\
\llap{$^{a}$}Special Research Centre for the Subatomic Structure of
Matter (CSSM), Department of Physics,
University of Adelaide, SA 5005, Australia
\\
\llap{$^{b}$}School of Physics, University of Edinburgh,
  Edinburgh EH9 3JZ, UK
\\
\llap{$^{c}$}Department of Physics, Zhejiang University, Hangzhou,
Zhejiang 310027, China
\\
E-mail: \\
\email{sboinepa@physics.adelaide.edu.au},

\email{jhedditc@physics.adelaide.edu.au},

\email{blasscoc@physics.adelaide.edu.au},

\email{dleinweb@physics.adelaide.edu.au},

\email{Anthony.Williams@adelaide.edu.au},

\email{jzanotti@ph.ed.ac.uk},

\email{jzhang@physics.adelaide.edu.au}}

\abstract{Fascinating aspects of the light quark-mass behavior of
  baryon electromagnetic form factors are highlighted.  Using FLIC
  fermions on $20^3 \times 40$ quenched ${\cal O}(a^2)$-improved gauge
  fields, we explore charge radii and magnetic moments at pion masses
  as light as 300 MeV.  Of particular interest is chiral curvature of
  proton charge radii and magnetic moments, the environmental
  dependence of strange quark properties in hyperons, and the
  remarkable signature of quenched chiral-nonanalytic behavior in the
  magnetic moment of $\Delta$ baryon resonances.}

\FullConference{
XXIV International Symposium on Lattice Field Theory\\
July 23-28 2006\\
Tucson Arizona, US}

\begin{document}

\section{Introduction}

With the advent of new improved fermion actions, it is now possible to
explore the light quark-mass regime of hadron electromagnetic form
factors on large physical volumes with unprecedented accuracy
\cite{Boinepalli:2006xd,mesonFF}.
In this report, we highlight a few of the most fascinating aspects of
the light quark-mass behavior of baryon electromagnetic structure.
For a complete discussion of these results and the associated lattice
techniques used to obtain them, we refer the interested reader to
Ref.~\cite{Boinepalli:2006xd}.  We also note that these calculations
have formed the foundation for precise determinations of strange-quark
contributions to proton electromagnetic form factors
\cite{Leinweber:2004tc,Leinweber:2005bz,Leinweber:2006ug}.

\section{Lattice Techniques}
   
The electromagnetic form factors are obtained using the three-point
function techniques established by Leinweber, {\it et al.} in
Refs.~\cite{Leinweber:1990dv,Leinweber:1992hy,Leinweber:1992pv} and
updated for smeared sources in Ref.~\cite{Boinepalli:2006xd}.
Our quenched gauge fields are generated with the ${\mathcal O}(a^2)$
mean-field improved Luscher-Weisz plaquette plus rectangle gauge
action \cite{Luscher:1984xn} using the plaquette measure for the mean
link.  400 quenched gauge field configurations on $20^3 \times 40$
lattices with lattice spacing $a= 0.128$ fm are generated via the
Cabibbo-Marinari pseudo-heat-bath algorithm~\cite{Cab82} using a
parallel algorithm with appropriate link partitioning
\cite{Bonnet:2000db}.

We use the fat-link irrelevant clover (FLIC) Dirac operator
\cite{Zanotti:2001yb} which provides a new form of nonperturbative
${\mathcal O}(a)$ improvement \cite{Zanotti:2004dr}.  The improved
chiral properties of FLIC fermions allow efficient access to the light
quark-mass regime \cite{Boinepalli:2004fz}, making them ideal for
dynamical fermion simulations now underway \cite{Kamleh:2004xk}.

Of particular interest, is our use of an ${\cal O}(a)$-improved FLIC
conserved vector current \cite{Boinepalli:2006xd}.  We follow the
technique proposed by Martinelli {\it et al.}  \cite{Martinelli:ny}.
The standard conserved vector current for Wilson-type fermions is
derived via the Noether procedure
\be
j_\mu^{\rm C} \equiv \frac{1}{4}\bigl[\overline{\psi}(x)\, (\gamma_\mu -
r)\, U_\mu(x)\, \psi(x+\hat{\mu}) 
+ \overline{\psi}(x+\hat{\mu})\, (\gamma_\mu + r)\, U_\mu^\dagger(x)\, 
\psi(x) 
+ (x\rightarrow x-\hat{\mu})\bigr] .
\label{conserved}
\ee
The ${\cal O}(a)$-improvement term is also derived from the fermion
action and is constructed in the form of a total four-divergence,
preserving charge conservation.  The ${\cal O}(a)$-improved conserved
vector current is
\be
j_\mu^{\rm CI} \equiv j_\mu^{\rm C} (x) + \frac{r}{2} C_{CVC}\, a \sum_\rho
\partial_\rho \bigl( \overline{\psi}(x) \sigma_{\rho\mu}\psi(x)\bigr)
\, ,
\label{impconserved}
\ee
where $C_{CVC}$ is the improvement coefficient for the conserved
vector current.

The terms proportional to the Wilson parameter $r$ in
Eq.~(\ref{conserved}) and the four-divergence in
Eq.~(\ref{impconserved}) have their origin in the irrelevant operators
of the fermion action and vanish in the continuum limit.
Non-perturbative improvement is achieved by constructing these terms
with fat-links.  Perturbative corrections are small
for fat-links and the use of the tree-level value for $C_{CVC} = 1$
together with small mean-field improvement corrections ensures that
${\cal O}(a)$ artifacts are accurately removed from the vector
current.  This is only possible when the current is constructed with
fat-links.  Otherwise, $C_{CVC}$ needs to be appropriately tuned to
ensure all ${\cal O}(a)$ artifacts are removed.

\section{Highlights}

A chief aim of the CSSM Lattice Collaboration has been to reveal the
electromagnetic structure of baryons near the chiral regime, search
for evidence of chiral nonanalytic curvature and examine the extent to
which the observed features are in accord with quenched chiral
effective field theory (\QchiPT) \cite{Leinweber:2002qb}.  In the
following we discuss some of the more interesting aspects of hadron
structure revealed in the light quark-mass regime.

\begin{figure}[tbp!]
\begin{center}
 {\includegraphics[height=0.65\hsize,angle=90]{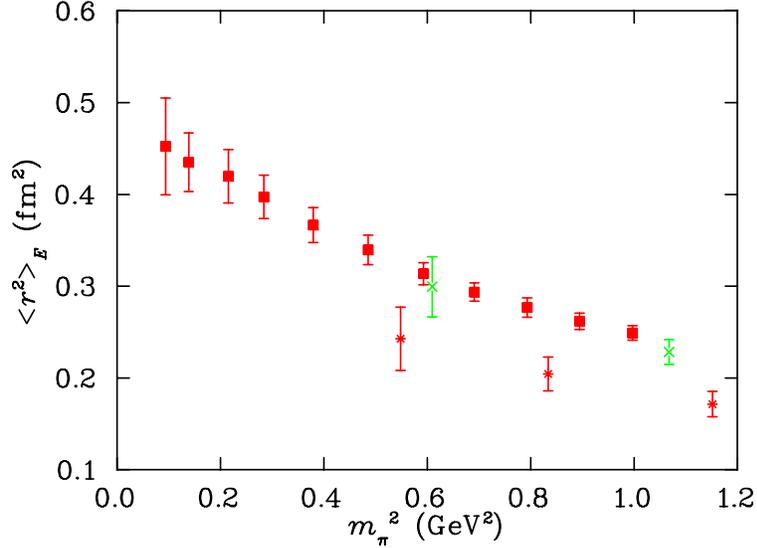}}
\end{center}
\vspace*{-0.5cm}
\caption{The proton charge radius is compared with previous state of
  the art lattice simulation results in quenched QCD.  The solid
  squares indicate current lattice QCD results with FLIC fermions.
  The stars indicate the lattice results of \cite{Leinweber:1990dv}
  while the crosses indicate the results of \cite{Wilcox:1991cq}, both
  of which use the standard Wilson actions for the gauge and fermion
  fields.}  
\vspace{-0.5cm}
\label{crp}
\end{figure}

\begin{figure}[tbp!]
\begin{center}
 {\includegraphics[height=0.65\hsize,angle=90]{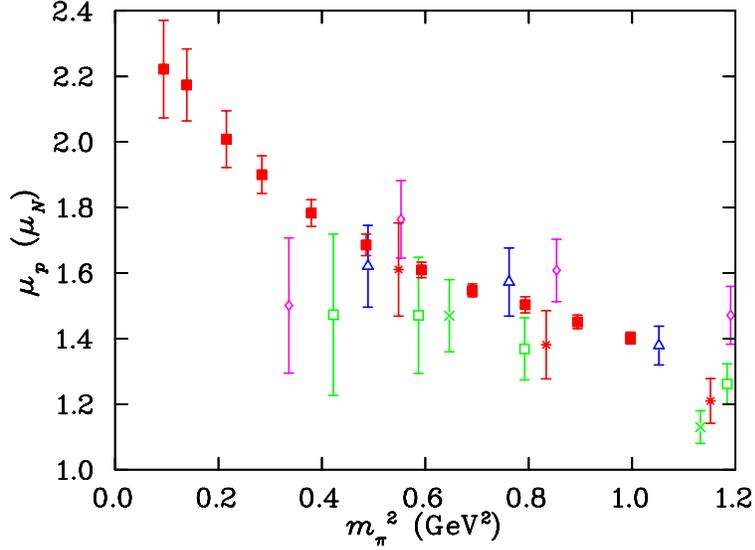}}
\end{center}
\vspace*{-0.5cm}
\caption{The proton magnetic moment in nuclear magnetons is compared
  with a variety of lattice simulations using three-point function
  techniques.  The solid squares indicate our current lattice QCD
  results with FLIC fermions.  The stars indicate the early lattice
  results of Ref.~\cite{Leinweber:1990dv}.  The crosses (only one
  point) indicate the results of Ref.~\cite{Wilcox:1991cq}.  The open
  symbols describe the QCDSF collaboration results
  \cite{Gockeler:2003ay}.  Open squares indicate results with
  $\beta=6.0$, open triangles indicate those with $\beta=6.2$ while
  the open diamonds indicate their results with $\beta=6.4$.}
\label{magmomprotcomp}
\end{figure}

We begin with an examination of the proton's charge radius and
magnetic moment.  In both cases, quenched chiral nonanalytic curvature
is predicted to act to increase the magnitude of these observables;
logarithmically for radii and as $m_q^{1/2}$ for magnetic moments.
Figures \ref{crp} and \ref{magmomprotcomp} illustrate our FLIC fermion
results in the context of other three-point function based lattice
calculations.  In both cases, curvature acting to increase the
magnitude of the results is observed.

Of particular interest is the environment sensitivity of quark sector
contributions to baryon electromagnetic properties, where some rather
interesting physics has been discovered.  Environment sensitivity is
easily observed in the strange quark sector contributions to hyperon
properties.  As the strange quark mass is held fixed while the light
quark masses are varied, any variation of the strange quark
contribution is a pure environmental effect.

Of course such effects are predicted in chiral effective field theory
where the mass of the Kaon changes as the light quark masses are
varied.  As the Kaon mass varies, the distribution of strange quarks
and their contributions to the baryon magnetic moment will change,
even when the strange quark mass is held constant.

\begin{figure}[tbp!]
\begin{center}
 {\includegraphics[height=0.65\hsize,angle=90]{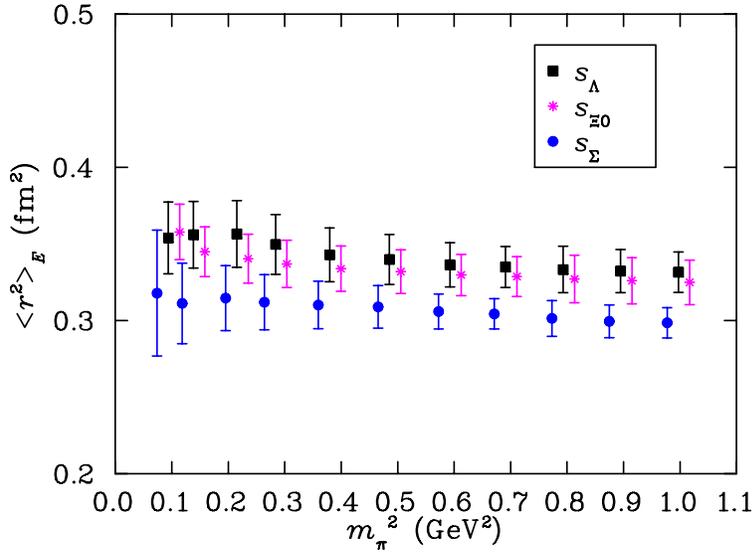}}
\end{center}
\vspace*{-0.5cm}
\caption{Electric charge distribution radii of strange quarks
  including ${s}_{\Lambda}$, ${s}_{\Xi^0}$ and ${s}_{\Sigma^0}$.  The
  data for ${s}_{\Xi^0}$ and ${s}_\Lambda$ are plotted at shifted
  ${m}^2_\pi$ values for clarity.}
\vspace{-0.2cm}
\label{crslxs}
\end{figure}

\begin{figure}[tbp!]
\begin{center}
 {\includegraphics[height=0.65\hsize,angle=90]{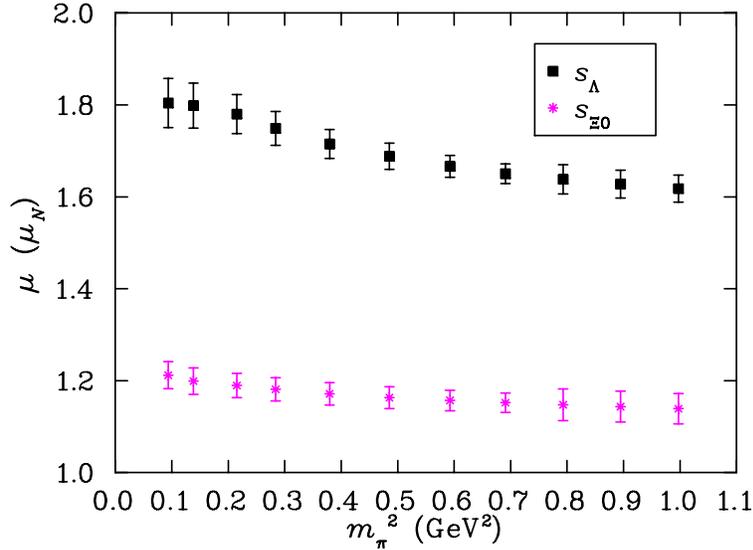}}
\end{center}
\vspace*{-0.5cm}
\caption{Magnetic moments of ${s}_{\Lambda}$ and ${s}_{\Xi^0}$ as a
  function of quark mass.}
\label{magmomslx}
\end{figure}

Figures~\ref{crslxs} and \ref{magmomslx} present results for charge
distribution radii and magnetic moments of strange quarks in hyperons
respectively.  These quark sector contributions are presented for a
single quark with unit charge.  The strange quark in $\Lambda$,
$s_\Lambda$, in an environment of two light quarks, displays the most
significant dependence on the light-quark sector.  It is interesting
that the coupling of $\Lambda$ to the energetically favoured $K N$
channel is large in \QchiPT\ \cite{Leinweber:2002qb}.  Moreover, the
sign of the chiral coefficient is such that the virtual transitions
act to enhance the charge distribution and magnetic moment as the
chiral limit is approached.
We have also confirmed a nontrivial role for light-quark contributions
to the Lambda magnetic moment
\cite{Boinepalli:2006xd,Leinweber:1990dv}.  In simple quark models the
light quarks are in an isospin and spin singlet configuration and do
not contribute.

\section{Quenched Chiral Artifacts}

While in many cases, the quenched approximation preserves the
qualitative features of full QCD, albeit with suppressed chiral
coefficients, there are some cases where the sickness of the quenched
approximation is fatal.  Perhaps the best known example is the $a_0$
meson correlator.  At sufficiently light quark masses, decays to the
negative-metric double-hairpin $\pi\, \eta'$ channel changes the sign
of the two-point function.

\begin{figure}[tb]
\begin{center}
{\includegraphics[height=0.65\hsize,angle=90]{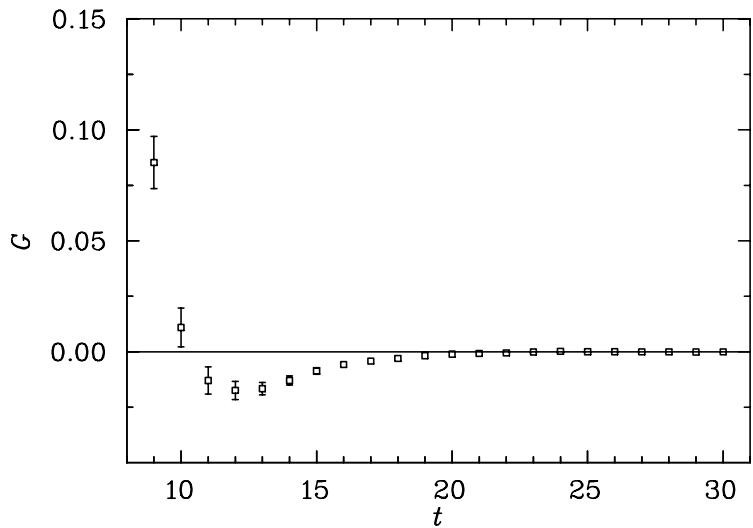}}
\caption{The signature of the quenched decay of the $a_0$ meson to the
   negative-metric $\pi\,\eta'$ channel.  The correlator becomes
   negative as the lowest-lying negative-metric $\pi\, \eta'$ decay
   channel saturates the correlator.}
\label{quencheda0}
\end{center}
\end{figure}

Figure \ref{quencheda0} displays this classic signature obtained with
FLIC fermions at our second lightest quark mass where $m_\pi =
372(6)$~MeV.  Whereas the correlator begins positive, it changes sign
as the lowest-lying negative-metric $\pi\, \eta'$ decay channel
saturates the correlator.

\begin{figure}[tbp!]
\begin{center}
  {\includegraphics[height=0.65\hsize,angle=90]{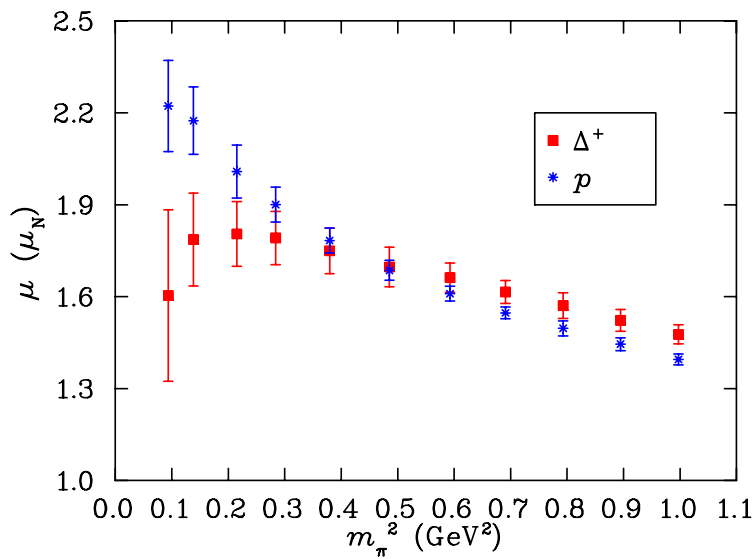}}
\end{center}
\caption{Magnetic moments of the $\Delta^+$ and the proton at quark
 masses where the $\Delta$ is stable by energy conservation.}
\label{magmomPD}
\end{figure}

A similarly dramatic signature, discovered by the CSSM Lattice
Collaboration in 2003 \cite{Leinweber:2003ux}, is displayed in the
magnetic moments of $\Delta$ baryons.  Figure~\ref{magmomPD} compares
the magnetic moment of the $\Delta^+$ with that calculated for the
proton on the same lattice.  The simplest quark model predicts that
the proton and the $\Delta^+$ have equal magnetic moments.  Upon
including hyperfine interactions the $\Delta^+$ moment is expected to
be larger and this is observed in Fig.~\ref{magmomPD} at large quark
masses.

However at light quark masses, one must turn attention to the meson
dressings of the baryons.  The presence of the $\Delta \to N \pi$
decay channel is particularly important for the quark-mass dependence of
$\Delta$ properties in general \cite{Leinweber:2003ux}.  Rapid
curvature associated with nonanalytic behavior is shifted to larger
pion masses near the $N$-$\Delta$ mass splitting, $m_\pi \sim M_\Delta
- M_N$.

In full QCD, pion-loop contributions are expected to enhance the
magnitude of both the proton and $\Delta^+$ magnetic moments
\cite{Cloet:2003jm}.  However a significant down turn is observed for
the $\Delta^+$ magnetic moment as the opening of the $N\, \pi$ decay
channel is approached.

As described in Ref.~\cite{Leinweber:2003ux}, this feature is in
accord with the expectations of \QchiPT.  Quenched-QCD decay-channel
contributions to $\Delta$ properties come with a sign opposite to that
of full QCD.  This artifact provides an unmistakable signature of the
quenched meson cloud.

Acknowledgment:  We thank the Australian Partnership for Advanced Computing (APAC) and the South Australian Partnership for Advanced Computing (SAPAC) for generous allocations of supercomputer time.


\begin{thebibliography}{99}

\bibitem{Boinepalli:2006xd}
  S.~Boinepalli, D.~B.~Leinweber, A.~G.~Williams, J.~M.~Zanotti and J.~B.~Zhang,
  {\it Precision electromagnetic structure of octet baryons in the chiral
  regime},
  arXiv:hep-lat/0604022.


\bibitem{mesonFF}
B. G. Lasscock, {\it et al.}, {\it Vector meson electromagnetic form
  factors}, these proceedings. 


\bibitem{Leinweber:2004tc}
  D.~B.~Leinweber {\it et al.},
  {\it Precise determination of the strangeness magnetic moment of the  nucleon},
  Phys.\ Rev.\ Lett.\  {\bf 94} (2005) 212001
  [arXiv:hep-lat/0406002].

\bibitem{Leinweber:2005bz}
  D.~B.~Leinweber, {\it et al.},
  {\it Systematic uncertainties in the precise determination of the strangeness
  magnetic moment of the nucleon},
  Eur.\ Phys.\ J.\ A {\bf 24S2} (2005) 79
  [arXiv:hep-lat/0502004].

\bibitem{Leinweber:2006ug}
  D.~B.~Leinweber {\it et al.},
  {\it Strange electric form factor of the proton},
  Phys.\ Rev.\ Lett.\ {\bf 97} (2006) 022001
  [arXiv:hep-lat/0601025].


\bibitem{Leinweber:1990dv}
  D.~B.~Leinweber, R.~M.~Woloshyn and T.~Draper,
  {\it Electromagnetic structure of octet baryons},
  Phys.\ Rev.\ D {\bf 43} (1991) 1659.


\bibitem{Leinweber:1992hy}
  D.~B.~Leinweber, T.~Draper and R.~M.~Woloshyn,
  {\it Decuplet baryon structure from lattice QCD},
  Phys.\ Rev.\ D {\bf 46} (1992) 3067
  [arXiv:hep-lat/9208025].


\bibitem{Leinweber:1992pv}
  D.~B.~Leinweber, T.~Draper and R.~M.~Woloshyn,
  {\it Baryon octet to decuplet electromagnetic transitions},
  Phys.\ Rev.\ D {\bf 48} (1993) 2230
  [arXiv:hep-lat/9212016].


\bibitem{Luscher:1984xn}
M.~Luscher and P.~Weisz,
{\it On-shell improved lattice gauge theories},
Commun.\ Math.\ Phys.\  {\bf 97} (1985) 59
[Erratum-ibid.\  {\bf 98} (1985) 433].


\bibitem{Cab82}
N. Cabibbo and E. Marinari, 
{\it A new method for updating SU(N) matrices in computer simulations of gauge theories},
Phys.\ Lett.\ {\bf B119} (1982) 387.

\bibitem{Bonnet:2000db}
F.~D.~Bonnet, D.~B.~Leinweber and A.~G.~Williams,
{\it General algorithm for improved lattice actions on parallel computing  architectures},
J.\ Comput.\ Phys.\  {\bf 170} (2001) 1
[arXiv:hep-lat/0001017].


\bibitem{Zanotti:2001yb}
J.~M.~Zanotti {\it et al.}  [CSSM Lattice Collaboration],
{\it Hadron masses from novel fat-link fermion actions},
Phys.\ Rev.\ D {\bf 65} (2002) 074507
[arXiv:hep-lat/0110216].

\bibitem{Zanotti:2004dr}
J.~M.~Zanotti, B.~Lasscock, D.~B.~Leinweber and A.~G.~Williams,
{\it Scaling of FLIC fermions},
Phys.\ Rev.\ D {\bf 71} (2005) 034510 
[arXiv:hep-lat/0405015].

\bibitem{Boinepalli:2004fz}
S.~Boinepalli, W.~Kamleh, D.~B.~Leinweber, A.~G.~Williams and J.~M.~Zanotti,
{\it Improved chiral properties of FLIC fermions},
Phys.\ Lett.\ B {\bf 616} (2005) 196 
[arXiv:hep-lat/0405026].

\bibitem{Kamleh:2004xk}
W.~Kamleh, D.~B.~Leinweber and A.~G.~Williams,
{\it Hybrid Monte Carlo with fat link fermion actions},
Phys.\ Rev.\ D {\bf 70} (2004) 014502
[arXiv:hep-lat/0403019].


\bibitem{Martinelli:ny}
G.~Martinelli, C.~T.~Sachrajda and A.~Vladikas,
{\it A study of 'improvement' in lattice QCD},
Nucl.\ Phys.\ B {\bf 358} (1991) 212.


\bibitem{Leinweber:2002qb}
  D.~B.~Leinweber,
  {\it Quark contributions to baryon magnetic moments in full, quenched and
  partially quenched QCD},
  Phys.\ Rev.\ D {\bf 69} (2004) 014005
  [arXiv:hep-lat/0211017].


\bibitem{Wilcox:1991cq}
  W.~Wilcox, T.~Draper and K.~F.~Liu,
  {\it Chiral limit of nucleon lattice electromagnetic form-factors},
  Phys.\ Rev.\ D {\bf 46} (1992) 1109
  [arXiv:hep-lat/9205015].


\bibitem{Gockeler:2003ay}
  M.~Gockeler {\it et al.},
                  [QCDSF Collaboration],
  {\it Nucleon electromagnetic form factors on the lattice and in chiral
  effective field theory},
  Phys.\ Rev.\ D {\bf 71} (2005) 034508
  [arXiv:hep-lat/0303019].


\bibitem{Leinweber:2003ux}
  D.~B.~Leinweber, A.~W.~Thomas, A.~G.~Williams, R.~D.~Young, J.~M.~Zanotti and J.~B.~Zhang,
  {\it Observing chiral nonanalytic behavior with FLIC fermions},
  Nucl.\ Phys.\ A {\bf 737} (2004) 177
  [arXiv:nucl-th/0308083].


\bibitem{Cloet:2003jm}
  I.~C.~Cloet, D.~B.~Leinweber and A.~W.~Thomas,
  {\it Delta baryon magnetic moments from lattice QCD},
  Phys.\ Lett.\ B {\bf 563} (2003) 157
  [arXiv:hep-lat/0302008].


\bibitem{Labrenz:1996jy}
J.~N.~Labrenz and S.~R.~Sharpe,
{\it Quenched chiral perturbation theory for baryons},
Phys.\ Rev.\ D {\bf 54} (1996) 4595
[arXiv:hep-lat/9605034].


\end{thebibliography}
\end{document}